\documentclass[12pt,tightenlines,eqsecnum,floats,aps,amsmath,amssymb,nofootinbib,prd,showpacs]{revtex4}
\usepackage{setspace} \usepackage{amsmath,amssymb,amsfonts,amsthm}
\usepackage{graphicx}

\newcommand{\be}{\begin{equation}}
\newcommand{\ee}{\end{equation}}
\newcommand{\beq}{\begin{eqnarray}}
\newcommand{\eeq}{\end{eqnarray}}

\begin{document}
\title{Tachyonic decay of unstable Dirichlet branes} \author{William
  Nelson\footnote{william.nelson@kcl.ac.uk}, Mairi
  Sakellariadou\footnote{mairi.sakellariadou@kcl.ac.uk}}
\affiliation{Department of Physics, King's College, University of
  London, Strand WC2R 2LS, London, U.K.}

\begin{abstract}
\noindent
We consider an unstable bound system of two supersymmetric Dirichlet
branes of different dimensionality ($p,p'$ with $p'<p$) embedded in a
flat non-compactified IIA or IIB type background. We study the decay,
via tachyonic condensation, of such unstable bound states leading to a
pair of bound D$(p-1)$, D$p'$-branes.  We show that only when the
gauge fields carried by the D$p$-brane are localised perependicular to
the tachyon direction, then tachyon condensation will indeed take
place.  We perform our analysis by combining both, the Hamiltonian and
the Lagrangian approach.

\end{abstract}
\vspace{.2cm}


\maketitle

\section{Introduction}

In the context of type II string theories, BPS Dirichlet branes with
even (within type IIA string theory) or odd (within type IIB) number
of tangential spatial directions can arise. Such objects, called
D$p$-branes, are invariant under half of the space-time supersymmetry
transformations of the theory and are charged under a $(p+1)$-form
gauge field, having its origin in the Ramond-Ramond
sector. Anti-branes ($\bar{\rm{D}}$) carry opposite Ramond-Ramond
charge and have opposite orientation than D-branes. Starting with a
D$p$-$\bar{\rm{D}}p$-brane (with $p$ even) in type IIA string theory,
and modding it out by the exact symmetry of the theory $(-1)^{F_L}$
(where $F_L$ denotes the contribution to the space-time fermion number
from the left-moving sector of the string world-sheet), one can
define~\cite{sen1998} a single non-BPS Dirichlet brane of type
IIB. Similarly, one can construct non-BPS Dirichlet branes of type
IIA. Thus, type IIB string theory contains BPS D-branes of odd
dimension and non-BPS D-branes of even dimension; the vice versa holds
for type IIA string theory. Alternatively, non-BPS Dirichlet branes
can be constructed as tachyonic kink solutions on the brane-anti-brane
system, with the energy density concentrated around a
$(p-1)$-dimensional space. One can thus claim an equivalence between
tachyonic kink solution and non-BPS Dirichlet branes.

In either type IIA or type IIB string theory, non-BPS Dirichlet branes
are unstable due to the appearance of a tachyonic mode. More
precisely, when the tachyon condenses to its minimum, the tachyonic
ground state cannot be distinguished from the vacuum, in the sense
that it carries neither charge nor energy density. Even though these
branes are unstable, one may obtain stable non-BPS Dirichlet branes by
projecting out the tachyonic mode.

In what follows, we study the decay, via tachyonic condensation, of
unstable bound states of Dirichlet (D) branes of different
dimensionality, embedded in a flat non-compactified $d$-dimensional
space-time. The choice of the background as a flat IIA, or IIB, type
will specify whether $p$ is even, or odd, respectively.  Our starting
point is Sen's effective action for unstable Dirichlet
branes~\cite{sen98}, which is presented in
Section~\ref{section:action}.  The decay of the branes becomes
apparent from the presence of a tachyonic mode on the brane
world-volume. To make our analysis more transparent we first discuss
the bosonic sector of the theory in Section~\ref{section:bosonic}, and
then the fermionic sector in Section~\ref{section:fermionic}. We
consider two complementary approaches, the Hamiltonian approach and
the Lagrangian one, following Refs.~\cite{Gibbons:2000hf} and
\cite{Sen:2003tm}, respectively.  In the first
approach~\cite{Gibbons:2000hf}, discussed in
Section~\ref{subsection:hamiltonian}, we express the Hamiltonian of
the system in terms of the canonical variables. Despite the fact that
in the tachyon condensation limit the Lagrangian vanishes, the
canonical momenta, and thus the Hamiltonian, remain well-defined.  In
the second approach~\cite{Sen:2003tm}, discussed in
Section~\ref{subsection:lagrangian}, the action remaining after
condensation is directly calculated. These two approaches have been so
far considered independently in the literature. However, they can work
together allowing us to understand the geometric restrictions that are
required for the system to fully condense. This is indeed the novelty
of our work. We discuss our findings in
Section~\ref{section:conclusions}.

We note that throughout this work we are using units in which the
fundamental string tension $g_{\rm s}$ is equal to $1/(2\pi)$.  The
world-volume signature is taken to be $\eta=(-+\cdots +)$.

\section{Tachyon effective action}
\label{section:action}

The classical tachyon effective Lagrangian describing the dynamics of
the tachyon field of a D$p$-brane (i.e., unstable Dirichlet brane) of
type IIA or IIB superstring theory is given by the sum of
Dirac-Born-Infeld (${\cal L}_{\rm DBI}$) and Wess-Zumino (${\cal
  L}_{\rm WZ}$) type terms, as
\be
{\cal L}_{\rm eff}={\cal L}_{\rm DBI}+{\cal L}_{\rm WZ}~.  
\ee 
To write down the effective Lagrangian for D$p$-branes, one considers
the local symmetries of such theories, which consist of a general
coordinate invariance of the world-volume, a local fermionic symmetry
and a U(1) gauge invariance.  The local symmetries are then used to
make a gauge choice in which the unphysical degrees of freedom to
vanish.

The Dirac-Born-Infeld type term of the effective Lagrangian, which
provides a good description of our system under the assumption that
$T$ is large while its second and higher derivatives are small, can be
written as~\footnote{This action has been basically taken from
  Ref.~\cite{Aganagic:1996nn}, however we have treated the tachyon
  similarly to the $X^I$. Thus, we have included the tachyon kinetic
  term inside the determinant, as discussed in
  Refs.~\cite{Garousi:2000tr,Bergshoeff:2000dq}.}  
\be\label{eq:action1} 
{\cal L}_{\rm DBI} = -V\left(T\right)\sqrt{ -\det \left(
  \Pi^I_\mu\Pi_{\nu I} + {\cal F}_{\mu\nu} +\partial_\mu T
  \partial_\nu T\right) }~, 
\ee 
where $\mu,\nu\dots = 0,1,\dots, p$ are the world-volume indices and
$I,J,\dots = 0,1,\dots,d-1$ are the target space indices; $T$ denotes
the tachyon field; $V(T)$ is the tachyon potential, identified with
the vanishing tension of the decaying branes. Note that we have
explicitly included the kinetic term as it plays a crucial r\^ole in
what follows.  

The supersymmetric quantity $\Pi^I_\mu$ is related to
another supersymmetric quantity $\partial_\mu\Theta$ through
\be
\Pi^I_\mu = \partial_\mu X^I - \bar{\Theta}\Gamma^I
\partial_\mu\Theta~, 
\ee
while the supersymmetric expression of the world-volume gauge field is
\beq
 {\cal F}_{\mu\nu} &=& F_{\mu\nu} - \bar{\Theta}\Psi
 \Gamma_I\partial_\mu\Theta\left( \partial_\nu X^I -
 \frac{1}{2}\bar{\Theta}\Gamma^I \partial_\nu\Theta\right) \nonumber
 \\ && + \bar{\Theta}\Psi\Gamma_I\partial_\nu\Theta\left(
 \partial_\mu X^I - \frac{1}{2}\bar{\Theta}\Gamma^I
 \partial_\mu\Theta\right)~, 
\label{WZ-L}
\eeq 
where $F_{\mu\nu}=\partial_\mu A_\nu -\partial_\nu A_\mu$ is the
world-volume electromagnetic field, and $X^I$ are world-volume scalars
that give the transverse motion of the brane. For a 10-dimensional
space-time, $\Theta$ are 32 component fermionic spinors and $\Gamma^I$
are the $32\times 32$ Dirac matrices.  Finally, $\Psi$ stands for
either $\Gamma_{11}=\Gamma_0\Gamma_1\cdots\Gamma_9$, if $p$ is even
(flat type IIA background), or for the Pauli matrix $\tau_3$, if $p$
is odd (flat type IIB background).

We impose the {\sl static gauge} condition on the transverse
scalars~\cite{Aganagic:1996nn},
\be 
X^I = \left\{ \begin{array}{cc} \sigma^\mu & \ \ \mbox{for}\ \ I< p+1 \\ 
X^m & \ \ \mbox{for}\ \ I \geq
  p+1 \end{array}~,\right.
\ee 
where $m,n\dots = p+1, p+2,\dots,d-1$ and $\sigma^\mu$ are the
world-volume coordinates.  In the static gauge condition, the $(p+1)$
target-space coordinates are identified with the world-volume
coordinates.  The remaining spatial coordinates can be seen as
transverse excitations on the D$p$-brane.  Half of the 32 components
of the $\Theta$ coordinates can be eliminated by gauge fixing the
fermionic sector, so that one of the Majorana-Weyl spinors $\Theta_\pm$
equals zero, namely~\cite{Aganagic:1996nn}
\be 
\Theta = \left( \begin{array}{c} \Theta_+ \\ \Theta_- \end{array}
\right) = \left( \begin{array}{c} \Theta_+ \\ 0 \end{array} \right)~.
\ee 
We note that $\left( {\mathbb I} \pm \Gamma_{11} \right) \Theta =
2\Theta_\pm$. Thus, Eq.~(\ref{eq:action1})
becomes~\cite{Aganagic:1996nn}
\be
\label{eq:action2} 
{\cal L}_{\rm DBI} = -V\left(T\right) \sqrt{-\det \left( \eta_{\mu\nu} +
  \partial_\mu X^m \partial_\nu X_m + F_{\mu\nu} + B_{\mu\nu} +
  \partial_\mu T \partial_\nu T \right)}~, 
\ee 
where
\be\label{equation:B_munu} 
B_{\mu\nu} \equiv -2\bar{\Theta}_+ \left( \Gamma_\nu+
\Gamma_m\partial_\nu X^m \right) \partial_\mu \Theta_+ + \left(
\bar{\Theta}_+\Gamma_I \partial_\mu\Theta_+ \right)\left(
\bar{\Theta}_+ \Gamma^I \partial_\nu\Theta_+\right)~.  
\ee

The Wess-Zumino type term in the effective Lagrangian is a
$(p+1)$-form, describing the coupling of the Ramond-Ramond background
field strengths to the Dirichlet brane. Under a local fermionic
symmetry, ${\cal L}_{\rm WZ}$ vanishes under contraction with
$\Theta_-$.  Thus, the total effective Lagrangian is given by
Eq.~(\ref{eq:action2}), namely
\be
\label{eq:actiontotal} 
{\cal L}_{\rm eff} = -V\left(T\right) \sqrt{-\det \left( \eta_{\mu\nu} +
  \partial_\mu X^m \partial_\nu X_m + F_{\mu\nu} + B_{\mu\nu} +
  \partial_\mu T \partial_\nu T \right)}~, 
\ee 
with $B_{\mu\nu}$ as defined in Eq.~(\ref{equation:B_munu}).
 

\section{Bosonic sector}
\label{section:bosonic}

Let us first focus on the bosonic sector of the theory,
i.e. $B_{\mu\nu}=0$; we include $\Theta_+$ in
Section~\ref{section:fermionic}.  The form of the tachyonic potential can
contain topological obstructions that prevent the system from fully
decaying. In particular, for non-BPS branes the potential is such that
$V\left(T\rightarrow \pm \infty\right) = 0$, with $V(0)$ a
maximum~\cite{Sen:2003tm}. In this situation, a $(p-1)$-dimensional
kink defect forms, preventing the tachyon reaching its true vacuum at
all space-time coordinates. It is well-known that such a kink
precisely reproduces the dynamics of a D$(p-1)$-brane in the absence
of world-volume gauge fields.  Here we show that this result is also
true when the gauge fields are turned on, with some intuitive
restrictions on the geometry of the gauge fields and the tachyon.

A D$p$-brane can contain a D$p'$-brane (with $p'$ smaller than $p$),
forming a bound state. This bound state is just a D$p$-brane with the
electromagnetic charge of the D$p'$-brane dissolved onto its
world-volume. Such a bound system turns out to be particularly
relevant here, as it is precisely the system we are dealing with.  We
will first investigate this system using the Hamiltonian approach and
we we will then proceed with the Lagrangian one.


\subsection{Hamiltonian approach}
\label{subsection:hamiltonian}

If we neglect the fermionic components, it was shown in
Ref~\cite{Gibbons:2000hf} that the Hamiltonian obtained from the
gauge fixed Lagrangian, Eq.~(\ref{eq:action2}), can be written as
\beq\label{eq:ham0}
{\cal H} &=& \Bigl[ \left(P^m_{X}\right)^2 + \left(P_T\right)^2 + \left(
\partial_i X_m P^m_{X} + \partial_i T P_T + F_{ij}P^i_{A} \right)^2
\nonumber \\&& + \left( P^i_{A} \partial_i X_m\right)^2 + \left( P^i_{A}
\partial_i T\right)^2 +\left( P^i_{A}\right)^2 + V^2\left(T\right) \det
\left( h\right) \Bigr]^{1/2}~, 
\eeq
where 
\be
h_{ij} \equiv \delta_{ij} + \partial_i X^m \partial_j X_m + F_{ij} +
\partial_i T \partial_j T~,\nonumber
\ee
with $i,j,\dots = 1,2,\dots,p$ the spatial world-volume indices and
$P^i_A, P^m_X, P_T$ the canonical momenta of $A_i, X_m, T$,
respectively~\footnote{In Appendix~A we describe, this rather tedious
  derivation, including the fermionic degrees of freedom.}. 

If we consider the set of variables $\chi_N = \left( A_i,X_m,T\right)$
where $N = 1\dots d$, and define $\partial_M \chi_N = 0$ for $M>p$,
then the above Hamiltonian, Eq.~(\ref{eq:ham0}), can be written formally
as
\be\label{eq:ham} 
{\cal H} = \left[ \left( P_\chi^N\right)^2 + \left( F_{MN}
P_\chi^N\right)^2 +V^2\left(T\right) \det\left(h\right) \right]^{1/2}~,
\ee 
where $P_\chi^N$ are the canonical momenta of $\chi_N$ and $F_{MN} =
\partial_M \chi_N - \partial_N \chi_M$. Equation~(\ref{eq:ham}) turns
out to be a particularly useful way of expressing the Hamiltonian for
including fermionic degrees of freedom, as one can see from
Section~\ref{section:fermionic}.

At this point it is worth making the following remark, comparing our
analysis and our subsequently obtained results with that of
Ref.~\cite{Gibbons:2000hf}.  The kinetic terms of the tachyon were
neglected in Ref.~\cite{Gibbons:2000hf}. It was stated that
they could be included at any point, on the same footing as the
transverse scalars. This is of course true, but only valid until
the limit $T\rightarrow \pm\infty$ is taken. It is indeed the
non-zero spatial tachyon derivatives that result in the localisation
of the kink. This is the reason for which a {\it string gas} solution
was found in Ref.~\cite{Gibbons:2000hf} as the end point of the
tachyon condensation. In the following we show that
accounting for the tachyon kinetic term leads to the expected
localisation.

The usefulness of the Hamiltonian approach is that it allows the
$T\rightarrow \pm\infty$ limit to be taken whilst explicitly keeping
track of the canonical momenta of the gauge fields, which must be
conserved throughout the tachyon evolution. In the $T\rightarrow
\pm\infty$ limit, $V\left(T\right) \rightarrow 0$ and the Hamiltonian,
Eq.~(\ref{eq:ham}), becomes
\be\label{eq:ham2}
 {\cal H} = \left[ \left( P_\chi^N\right)^2 + \left( F_{MN}
 P_\chi^M\right)^2\right]^{1/2}~;
\ee
we have taken the $T\rightarrow \pm\infty$ limit in the canonical
momenta.

To show how this is derived, let us consider the specific case of a
three-dimensional non-BPS brane~\footnote{One can generalise this
  argument to other dimensionality, by applying T-duality.}.  In
this case, the tachyon is a function of only one world-volume
coordinate, which we will take, without loss of generality, to be the
$x$-direction.  Following the procedure outlined in
Ref.~\cite{Sen:2003tm}, we consider the field configuration
\be
 T(x) = f\left(ax\right)~,
\ee
where $f(w)$ is some arbitrary function, which however must satisfy
\be
f(-w)=-f(w)\ \ ,\ \ f'(w)>0\ \ \forall w\ \ ,\ \ f(\pm\infty)=\pm\infty~;
\ee
$a$ is a constant which we will later take to infinity. Thus,
$T=+\infty$ ($T=-\infty$), for $x>0$ ($x<0$).  Expanding
Eq.~(\ref{eq:ham2}) into terms that depend on the tachyon and its
derivatives and those that do not,
we obtain

\beq
 {\cal H} &=& \left[ \left( {\cal H}_{{\rm mod}\ T} \right)^2 + 
\left( P_\chi^i \partial_iT\right)^2 
 + \left( P_T \partial_iT\right)^2 + P_T^2 \right]^{1/2}~, \\
&=& \left[ \left( {\cal H}_{{\rm mod}\ T}\right)^2 + 
a^2\left[f'(ax)\right]^2\left( P_\chi^x\right)^2
+P_T^2\left\{a^2\left[ f'(ax)\right]^2 + 1\right\}\right]^{1/2}~;
\eeq
here $\chi$ represents the degrees of freedom {\it other} than the
tachyon and ${\cal H}_{{\rm mod}\ T}$ is the Hamiltonian of a
D$p$-brane (in this case $p=3$) without a tachyon. 

We are concerned with static solutions after the tachyon has fully
condensed, in which case $P_T=0$, since $\dot{T}=0$. Thus, for the
Hamiltonian to be finite in the limit $a\rightarrow \infty$, we
require $P_\chi^x=0$, i.e. the canonical momenta of all the
world-volume degrees of freedom are constrained to be perpendicular to
the direction of the kink. In this case, ${\cal H}_{{\rm mod}\ T}$ is
just the Hamiltonian of a D$(p-1)$-brane (in the case we consider
here, this is a two-dimensional Dirichlet brane) with no tachyon, as
indeed expected. In particular, we note that the canonical momentum of
the world-volume electromagnetic field is perpendicular to the kink
formed at the end of tachyon condensation and hence it is conserved.

We extend the above analysis by considering (not necessarily small)
fluctuations of bosonic fields around the kink background.  We study
the simple case of a translation along the $x$-direction, which
corresponds to fluctuations of the tachyon $T$ of the form
\be
\label{tachyon-fluct}
T = f\left(a\left(x-t(y,z)\right)\right)~.  
\ee 
In this case we find
\beq {\cal H} &=& \left[ \left( {\cal H}_{{\rm mod}\ T}\right)^2 +
  a^2\left[ f'(a(x-t(y,z)))\right]^2 \left\{ P_\chi^x +\left( P_\chi^y
  \partial_2 t\right)^2 + \left( P_\chi^z \partial_3t\right)^2
  \right\}^2 \right.  \nonumber\\ &&\left.\ \ + a^2\left[
    f'(a(x-t(y,z))) P_T\right]^2 \left\{ 1+ \left( \partial_y
  t\right)^2+ \left( \partial_z t\right)^2 \right\} +P_T^2
  \right]^{1/2}~.
\eeq
Looking for static solutions in which $P_T = 0$, we find again that
for the Hamiltonian to be finite in the $a\rightarrow \infty$ limit,
the canonical momenta of the degrees of freedom, $P_\chi$, must be
perpendicular to the direction of the kink. We are thus left again
with the Hamiltonian of a two-dimensional  Dirichlet brane.

We have thus shown that the Dirac-Born-Infeld (DBI) action describing the
dynamics of the tachyon field on a non-BPS D$p$-brane has a kink solution
described by the DBI action on a BPS D$(p-1)$-brane.


\subsection{Lagrangian approach}
\label{subsection:lagrangian}

Instead of working with the Hamiltonian, one can directly evaluate the
Lagrangian, Eq.~(\ref{eq:action2}). To do so, we follow the method
described in Ref.~\cite{Sen:2003tm} and extend it to explicitly
include non-zero world-volume electromagnetic fields, by using the fact that
the canonical momenta are constrained to lie perpendicular to the
tachyon. Without the previous Hamiltonian analysis this is an assumption,
however here it is a consistency requirement if the dynamics are to reach
a fully condensed configuration. The action given by Eq.~(\ref{eq:action2}) for a D3-brane,
in the absence of fermionic degrees of freedom, is just
\be
 S = -\int {\rm d}t {\rm d}x {\rm d} y {\rm d}z \ V\left(T\right)
  \sqrt{-\det {\cal M}}~,
\ee
where ${\cal M}_{\mu\nu} = \eta_{\mu\nu} + \partial_\mu X^m
\partial_\nu X_m + F_{\mu\nu} + \partial_\mu T \partial_\nu T$. 

We consider that the tachyon is given from the field configuration
described in Eq.~(\ref{tachyon-fluct}) and use that for the
Hamiltonian to be finite, all the degrees of freedom must evolve
perpendicular to the kink (the kink is along the $x$-direction and $y,z$
coordinates are perpendicular to $x$-direction), namely

\be 
A_\mu\left(x,t(y,z)\right) = \left\{ \begin{array}{cc} 0 &
  \ \ \mbox{for}\ \ \mu=x \\ a_\alpha\left(t(y,z)\right) &
  \ \ \mbox{for}\ \ \mu=\alpha \end{array}\right.  \ \ \ \ \ {\rm and}
\ \ \ \ \  X^I\left(x,t(y,z)\right) = x^I\left(t(y,z)\right)~;
\ee 
$A_\mu, X^I$ fields are independent of $x$. Thus, we obtain \beq {\cal
  M}_{xx} &=& 1+a^2\left(f'\right)^2~, \\
{\cal M}_{x \alpha} &=& {\cal M}_{ \alpha x} = -a^2\left( f'\right)^2
\partial_\alpha t~,\\
{\cal M}_{\alpha \beta} &=& \left[a^2 \left(f'\right)^2-1\right]
\partial_\alpha t \partial_\beta t + {\mathsf m}_{\alpha \beta}~, 
\eeq
where 
\be
{\mathsf m}_{\alpha \beta} \equiv \eta_{\alpha \beta} + \partial_\alpha x^m
\partial_\beta x_m + F_{\alpha\beta} + \partial_\alpha t
\partial_\beta t~, \ee
is just the equivalent of ${\cal M}$ with terms depending of the
$x$-direction being removed. 

By manipulating the determinant of the metric (as in Ref.~\cite{Sen:2003tm}),
to first order, we find
\be 
\det {\cal M} \sim a^2\left( f'\right)^2 \left( \det {\mathsf
  m}\right)~.  
\ee 
Thus, the action, in the limit $a\rightarrow \infty$, i.e. after the
tachyon condensation, is 
\beq
\label{actionD2}
\lim_{a \rightarrow \infty} S &=&-\int {\rm d}t\ {\rm d}x\ {\rm d}y
 \ {\rm
  d}z \ V\ \left(f\right)\ af'\ \sqrt{ -\det {\mathsf m}}~, \nonumber \\ 
&=& -{\cal T}_2
\int {\rm d}t\ {\rm d}y\ {\rm d}x \ \sqrt{-\det {\mathsf m}}~, 
\eeq 
where 
\be
{\cal T}_2 \equiv \int V(\tilde{x}){\rm d}\tilde{x}~,
\ee 
is taken to be the tension of the resulting D2-brane. Equation
(\ref{actionD2}b) is the world-volume action of a D2-brane.

Thus, the classical effective theory describing the dynamics of the
tachyon field on a non-BPS D$p$-brane has a kink solution of finite
tension described by a co-dimension one BPS brane.


\section{Fermionic Sector}\label{section:fermionic}

Including the spinors rapidly leads to rather complicated equations,
however these can be simplified by considering a change of (fermionic)
coordinates
\be
 \partial_\mu Y_I = \bar{\Theta}_+\Gamma_I \partial_\mu \Theta_+~,
\ee
with the restriction,
\be\label{eq:restrict}
 \partial_\mu X_n \left( \partial_\nu Y_m\right)^\dagger =
 \partial_\nu X_m \left( \partial_\mu Y_n \right)^\dagger~.
\ee
One can then explicitly check that the equations of motion derived
from
\beq\label{eq:action_fermi}
\hat{\cal L} &=& -V\left(T\right) \Big[ -\det \left( \eta_{\mu\nu} 
+ \partial_\mu X^m \partial_\nu X_m - 2 \partial_\mu Y_\nu 
- 2\partial_\mu Y^m \partial_\nu Y_m \right. \nonumber\\
&&{\hskip 2.7truecm}\left.+ 
\partial_\mu Y^I \partial_\nu Y_I + F_{\mu\nu} 
+ \partial_\mu T \partial_\nu T \right)\Big ]^{1/2}
\eeq
coincide with those obtained from Eq.~(\ref{eq:action2}). 

To be explicit, the equations of motion ({\sl e.o.m.}) obtained from
Eq.~(\ref{eq:action2}) are
\beq
\label{eq:EOM1}
 \partial_\nu X^I -\bar{\Theta}_+ \Gamma^I \partial_\nu \Theta_+ &=&
 0\ \ \ \ , \ \ \ \ \mbox{{\sl e.o.m.}\ \ for\ \ } \bar{\Theta}_+ \\
\label{eq:EOM2}
\left( \frac{1}{2}V\left(T\right) \sqrt{-\det{\cal M}} {\cal
  M}^{-1}_{[ \mu\nu]}\partial^\nu X_m \right)^{,\mu} &=& 0\ \ \ \ ,
\ \ \ \ \mbox{{\sl e.o.m.}\ \  for\ \ }  X  \\
\label{eq:EOM3}
\frac{1}{2}V\left(T\right)\sqrt{-\det{\cal M}} {\cal M}^{-1}_{[
    \mu\nu]}\partial^\mu X^m \left( \bar{\Theta}_+ \Gamma_m
\right)^{,\nu} &=& 0\ \ \ \ , \ \ \ \ 
\mbox{{\sl e.o.m.}\ \  for\ \ } \Theta_+~,
\eeq 
where 
${\cal M}_{\mu\nu} = \eta_{\mu\nu} + \partial_\mu X^m
\partial_\nu X_m + F_{\mu\nu} + B_{\mu\nu}+ \partial_\mu T
\partial_\nu T$. 
Equation~(\ref{eq:EOM3}) restricts, $\partial_\mu X^m \partial_\nu
\bar{\Theta}_+ \Gamma_m$ to be symmetric in $\mu$ and $\nu$.

The relevant equations of motion from Eq.~(\ref{eq:action_fermi}) are,
\beq
\label{eq:EOM_Y1}
\left( \frac{1}{2} V\left(T\right)\sqrt{-\det{\cal N}} {\cal
  N}^{-1}_{\mu\nu}\left( \partial^\nu X_m - 2 \partial^\mu Y_m
\right)\right)^{,\nu} + \left( \frac{1}{2} V\left(T\right)
\sqrt{-\det{\cal N}} {\cal N}^{-1}_{\mu\nu}\partial^\nu X_m
\right)^{,\mu} &&
\nonumber \\ \nonumber\\ 
&&{\hskip-5.5truecm}=0
\ \ \ \ ,
\ \ \ \ \mbox{{\sl e.o.m.}\ \  for\ \ }  X
\\  \nonumber\\  \nonumber\\ 
\label{eq:EOM_Y2}
\left( \frac{1}{2} V\left(T\right) \sqrt{ -\det{\cal N}} {\cal
  N}^{-1}_{\mu\nu} \left( \partial^\nu Y_m -2\partial^\nu X_m
\right)\right)^{,\mu} + \left( \frac{1}{2} V\left(T\right)
\sqrt{-\det{\cal N}} {\cal N}^{-1}_{\mu\nu} \partial^\mu Y_m
\right)^{,\beta} && \nonumber\\ \nonumber\\ &&{\hskip-5.5truecm}= 0
\ \ \ \ , \ \ \ \ \mbox{{\sl e.o.m.}\ \ for\ \ } Y_m
\\ \nonumber\\ \nonumber\\
\label{eq:EOM_Y3}
\left(\frac{1}{2} V\left(T\right) \sqrt{-\det{\cal N}}{\cal
  N}^{-1}_{\mu\nu} \left( \partial^\nu Y_\gamma -2\delta^\mu_\gamma
\right)\right)^{,\mu} +\left( \frac{1}{2} V\left(T\right)
\sqrt{-\det{\cal N}}{\cal N}^{-1}_{\mu\nu} \partial^\mu Y_\gamma
\right) ^{,\nu} 
&&
\nonumber \\ \nonumber\\
&&{\hskip-5.5truecm}=
0
\ \ \ \ ,
\ \ \ \ \mbox{{\sl e.o.m.}\ \  for\ \ }  Y_\gamma~, 
\eeq 
where 
\be
{\cal N}_{\mu\nu} = \eta_{\mu\nu} + \partial_\mu X^m \partial_\nu X_m
+ F_{\mu\nu} - 2 \partial_\mu Y_\nu - 2\partial_\mu Y^m \partial_\nu
Y_m + \partial_\mu Y^I \partial_\nu Y_I + \partial_\mu T \partial_\nu
T~.
\ee 
Equations (\ref{eq:EOM_Y1}) and (\ref{eq:EOM_Y2}) imply that
$\partial_\mu X_m = \partial_\nu Y_m$. Setting $\partial_\mu Y_\nu =
\eta_{\mu\nu}$, to be compatible with the static gauge (this can be
seen by doing this change of variables before enforcing the static
gauge constraint), one can recover Eq.~(\ref{eq:EOM1}).  Thus,
Eq.~(\ref{eq:EOM_Y2}) reproduces Eq.~(\ref{eq:EOM2}). By doing so, we
 we have lost the equation of motion for $\Theta_+$,
Eq.~(\ref{eq:EOM3}), however the restriction we place on $\partial_\mu
Y_m$, Eq.~(\ref{eq:restrict}), ensures that Eq.~(\ref{eq:EOM3}) is
always satisfied. In this sense, the Lagrangian given in
Eq.~(\ref{eq:action_fermi}) produces the {\it on shell} dynamics of
the brane. We finally make the remark that the equations of motion for
the tachyon and world-volume gauge fields are the same.

In these coordinates, the Hamiltonian becomes (see, Appendix~A)
\beq\label{eq:ham_Y}
 {\cal H} &=& \Bigl[ \left( P_X^m\right)^2 +\left(P_T\right)^2
   +\left(P_Y^I\right)^2 + \left(I_i\right)^2 + \left(P_A^i\partial_i
   X^m\right)^2 \nonumber \\ && \left(P_A^i \partial_iT\right)^2 +
   \left( P_A^i \partial_i Y_I\right)^2 + \left(P_A^i\right)^2 +
   V\left(T\right)^2 \det\left(h\right) \Bigr]^{1/2}~.
\eeq
This is precisely the Hamiltonian one arrives at by using the
Lagrangian Eq.~(\ref{eq:action_fermi}), and the result of the equation
of motion, $\partial_\mu Y^I = \partial_\mu X^I$, since in that case
the new coordinates, $Y_I$, appear on the same footing as the
transverse scalars, $X_I$, which is how they enter
Eq.~(\ref{eq:ham_Y}).

Finally, we note that just as in Ref.~\cite{Gibbons:2000hf}, this
Hamiltonian can formally be written as
\be
{\cal H} = \sqrt{ \left(P_\chi^N \right)^2 + \left( P_\chi^M
  F_{MN}\right)^2 }~,
\ee
where $\chi_N$ is the set of variables, $\left( A_i, X_m, T,
Y_I\right)$ with $N,M\dots = 1,2,\dots,d+d$ and $P_\chi^N$ are their
canonical momenta, with $F_{MN} = \partial_M \chi_N - \partial_N
\chi_M$, where
\be
 \partial_N = \left\{\begin{array}{cc}
                  \partial_M  & {\rm for} \ \ N \leq p \\
                   0 & {\rm for} \ \ N > p \end{array} \right.~.
\ee
Thus the conclusions of the bosnic case carry over immediately for the
explicit inclusion of fermionic degrees of freedom.


\section{Conclusions}
\label{section:conclusions}

We have shown that a non-BPS D$p$-brane carrying non-zero world volume
electromagnetic fields decays into a D$(p-1)$-brane with the
electromagnetic fields conserved and localised on that brane. The
requirement that these gauge fields be conserved manifests itself as a
restriction on the direction along which the tachyon can decay. 

In previous literature, it was assumed that any gauge fields on the
D$p$-brane were perpendicular to the tachyon, enabling the
D$(p-1)$-brane to form. Here, we have used the Hamiltonian approach,
to demonstrate that it is only when this condition is met that the
tachyon can indeed condense.

We have considered that the tachyon depends on a particular direction
and shown that only when the gauge fields are localised perpendicular
to this direction, is it possible for the tachyon condensation to
occur. A corollary of course is that if the gauge fields are non-zero
in all the world volume directions, then tachyon condensation cannot
occur, or at least the energy produced by coupling between the
tachyon's spatial derivatives and the gauge fields would prevent the
true tachyon vacuum by being reached.

In particular, this shows that a bound system composed by two branes
of different dimensionality (D$p$, D$p'$ bound brane system with
$p'<p$), which can be described by just a D$p$-brane with world-volume
gauge fields, decays exactly as a standard D$p$-brane, namely by
forming a bound state of a D$(p-1)$ and a D$p'$-brane.

Finally, it would be interesting to investigate whether we can recover
these results by following the gauge theory approach of
brane-antibrane systems discussed in Ref.~\cite{Gava:1997jt}.


\appendix
\label{app}
\section{}

We start by extracting the time derivatives of the Lagrangian
Eq.~(\ref{eq:action2}) using the identity~\cite{Gibbons:2000hf},
\beq
&& \det \left({\cal M_{\mu\nu} }\right) \equiv \det\left(
\eta_{\mu\nu} + \partial_\mu X^m \partial_\nu X_m + F_{\mu\nu} +
B_{\mu\nu} + \partial_\mu T\partial_\nu T\right) \nonumber \\ 
&&{\hskip1.8truecm} =
\left( \dot{X}^m\dot{X}_m + \dot{T}\dot{T} + B_{00} -1 \right)
\det\left(h\right) + E^{(+)}_i D_{ij}\left(h\right)E^{(-)}_j~, \eeq
where 
$$E^{(\pm)}_i = F_{0i} \pm \dot{X}^m \partial_i X_m\pm B_{0i} \pm
\dot{T}\partial_i T~,$$
$$h_{ij} = \delta_{ij} + \partial_i X^m
\partial_j X_m + F_{ij} + B_{ij} + \partial_i T \partial_j T~,$$ 
and
$$D_{ij}\left(h\right) - \left(-1\right)^{i+j}
\Delta_{ij}\left(h\right)~$$
with $\Delta_{ij}\left(h\right)$ being the determinant of $h$ with the
i$^{\rm th}$ row and j$^{\rm th}$ column removed. Note that, when the
inverse exists, we have $D_{ij}\left(h\right) =
\det\left(h\right)h^{-1}_{ij}$. Here we have used the fact that
$B_{ij} = B_{ji}$, by employing the equation of motion for
$\bar{\Theta}_+$.

Using this we find that the canonical momenta are
\beq
 P_X^m &=& \frac{-V\left(T\right)}{\sqrt{-\det {\cal M}}} \left[
   -\dot{X}^m\det\left(h\right) +\frac{1}{2}\left(
   E^{(+)}_iD_{ij}\left(h\right) - D_{ji} R^{(-)}_i \right) \partial_j
   X^m\right]~,\\ \nonumber\\ \nonumber\\
P_{\Theta_+} &=& \frac{-V\left(T\right)}{\sqrt{-\det {\cal M}}} \Bigl[
  \left( \bar{\Theta}_+ \Gamma_I \dot{\Theta}_+
  \right)\left(\bar{\Theta}_+ \Gamma^I \right) \det \left( h\right)
  \nonumber \\ &&{\hskip-.8truecm}+\frac{1}{2}\left( \bar{\Theta}_+
  \Gamma_I\right) \left( \bar{\Theta}_+\Gamma^I \partial_i \Theta_+
  \right) D_{ij} \left(h\right) E^{(-)}_j - \frac{1}{2}E^{(+)}_i
  D_{ij}\left(h\right)\left( \bar{\Theta}_+\Gamma_I\partial_j\Theta_+
  \right)\left( \bar{\Theta}_+\Gamma^I\right) \Bigr]~, \\ \nonumber
\\ \nonumber\\
P_A^i &=& \frac{V\left(T\right)}{2\sqrt{-\det{\cal M}}} \left[
  D_{ij}\left(h\right)E^{(-)}_j + E^{(+)}_j D_{ji}\left(h\right)
  \right]~, \\ \nonumber\\ \nonumber\\
P_T &=& \frac{-V\left(T\right)}{\sqrt{-\det {\cal M}}} \left[
  -\dot{T}\det\left(h\right) + \frac{1}{2}\left( E^{(+)}_i
  D_{ij}\left(h\right) - D_{ij}\left(h\right) E^{(-)}_j\right)
  \partial_j T \right]~,
\eeq
where we have used the equation of motion for $\bar{\Theta}$ to write
$$B_{\mu\nu} = -\left( \bar{\Theta}_+
\Gamma_I\partial_\mu\Theta_+\right)\left(\bar{\Theta}_+\Gamma^I
\partial_\nu\Theta_+\right)$$ and assumed $V\left(T\right) \neq
V\left(T,\dot{T}\right)$. Thus, the Hamiltonian can be written as
\be
 {\cal H} = P_X^I \dot{X}_I + P_{\Theta_+}\dot{\Theta}_+ + P_A^i
 F_{0i} + P_T \dot{T} - {\cal L} =
 \frac{V\left(T\right)\det\left(h\right)}{\sqrt{-\det {\cal M}}}~, 
\ee
where we have fixed the final gauge freedom via $\partial_i A_0 =0$.

Let us define
\be
C_I \equiv {\cal H} \left[ \left(\bar{\Theta}_+\Gamma_I
  \dot{\Theta}_+\right) -\frac{1}{2}\left( E^{(+)}_j h^{-1}_{ji} -
  h^{-1}_{ij}E^{(-)}_j\right)\left(\bar{\Theta}_+\Gamma_I\partial_i\Theta_+
  \right)\right]~,
\ee
and use the identities
\beq
\frac{\cal H}{2}\left( E^{(+)}_j h^{-1}_{ji} - E^{(-)}_j
h^{-1}_{ij}\right) &=& \partial_iX_m P_X^m + \partial_iT P_T + F_{ij}
P^i_A + \partial_i\Theta_+ P_{\Theta_+} \equiv I_i~, 
\nonumber \\ \\ \nonumber\\
{\cal H}^2 \left[ \dot{X}_m\dot{X}^m + \dot{T}\dot{T} + B_{00} +
  E^{(+)}_j h^{-1}_{ji} E^{(-)}_i \right] &=& \left(P_X^m\right)^2 +
\left(P_T\right)^2 + \left(C_I\right)^2 + \left( I_i\right)^2 +
\left(P_A^i\partial_iX^m\right)^2 \nonumber \\ &&
{\hskip-1.8truecm}
+ \left( P_A^i
\partial_i T\right)^2 + \left[ P_A^i \left( \bar{\Theta}_+\Gamma_I
  \partial_i\Theta_+\right)\right]^2 +\left(P_A^i\right)^2~, 
\\ \nonumber \\ \nonumber\\
{\cal H}^2\left[ 1-\dot{X}_m\dot{X}^m -\dot{T}\dot{T} -B_{00} 0 E^{(+)}_i
h^{-1}_{ij}E^{(-)}_j\right] &=& V\left(T\right)^2\det\left(h\right)~,
\eeq
to find
\beq\label{eq:ham_app}
 {\cal H} &=& \Bigl[ \left(P_X^m\right)^2 + \left( P_T\right)^2
   +\left(C_I\right)^2+ \left( \partial_i X_m P_X^m + \partial_iT P_T
   + F_{ij}P^j_A + \partial_i \Theta_+ P_{\Theta_+} \right)^2
   \nonumber \\ &&{\hskip-1.8truecm}+ \left(P_A^i \partial_iX^m\right)^2 +\left(
   P_A^i\partial_iT\right)^2 +\left( P_A^i \bar{\Theta}_+ \Gamma_I
   \partial_i \Theta_+\right)^2 + \left(P_A^i\right)^2 +
   V\left(T\right)^2 \det\left(h\right) \Bigr]^{1/2}~.
\eeq 
Notice that the bosonic case, $\Theta_+ = \bar{\Theta}_+ =
P_{\Theta_+} = C_I = 0$ gives us Eq.~(\ref{eq:ham0}), as expected.

The fermionic terms lend themselves to a simplification, by noting that
\be
 P_{\Theta_+} = C_I \bar{\Theta}_+\Gamma^I~.
\ee
Thus, we see that all the $\Theta_+$ terms occur in the combination
\be
 \bar{\Theta}_+ \Gamma_I \partial_i \Theta_+~, \ \ \ {\rm and} \ \ \ C_I~.
\ee
If we define new coordinates $Y_I$, such that $\partial_\mu Y_I =
\bar{\Theta}_+\Gamma_I \partial_\mu\Theta_+$, then
\be
 P_Y^I = {\cal H}\left[ \left( \bar{\Theta}_+\Gamma^I\dot{\Theta}_+ \right)+
-\frac{1}{2}\left( E^{(+)}_ih^{-1}_{ij} - h^{-1}_{ij}E^{(-)}_j\right)
\left( \bar{\Theta}_+\Gamma^I \partial_i\Theta_+\right)\right] = C^I~.
\ee
So, in terms of $Y_I$ the Hamiltonian Eq~(\ref{eq:ham_app}) becomes
\beq
 {\cal H} &=& \Bigl[ \left( P_X^m\right)^2 +\left(P_T\right)^2
   +\left(P_Y^I\right)^2 + \left(I_i\right)^2 + \left(P_A^i\partial_i
   X^m\right)^2 \nonumber \\ && \left(P_A^i \partial_iT\right)^2 +
   \left( P_A^i \partial_i Y_I\right)^2 + \left(P_A^i\right)^2 +
   V\left(T\right)^2 \det\left(h\right) \Bigr]^{1/2}~.
\eeq

\acknowledgments It is a pleasure to thank Riccardo Sturani for
collaboration during early stages of this work.  This work is
partially supported by the European Union through the Marie Curie
Research and Training Network {\sl UniverseNet} (MRTN-CT-2006-035863).

\end{document}